\documentclass[12pt]{article}
\usepackage{graphicx}
\usepackage{amsmath}
\usepackage{amssymb}
\usepackage{caption2}
\begin{document}
\bibliographystyle{prsty}
\begin{center}
{\large {\bf \sc{   $D_{s0}(2317)$ as a tetraquark state with QCD sum rules in heavy quark limit }}} \\[2mm]
Zhi-Gang Wang$^{1}$ \footnote{Corresponding author; E-mail,wangzgyiti@yahoo.com.cn.  } and Shao-Long Wan$^{2}$      \\
$^{1}$ Department of Physics, North China Electric Power University, Baoding 071003, P. R. China \\
$^{2}$ Department of Modern Physics, University of Science and Technology of China, Hefei 230026, P. R. China \\
\end{center}

\begin{abstract}
In this article, we take the point of view that the charmed scalar
meson $D_{s0}(2317)$ be a tetraquark state and devote to calculate
its mass within the framework of the QCD sum rules approach in the
heavy quark limit. The numerical values for the mass of the
$D_{s0}(2317)$ are consistent with the experimental data, there must
be some tetraquark component in the scalar meson $D_{s0}(2317)$.
Detailed discussions about the threshold parameter and Borel
parameter for the multiquark states are also presented.
\end{abstract}

 PACS number: 12.38.Aw, 12.38.Qk

Key words: $D_s(2317)$, QCD sum rules

\section{Introduction}
The  discovery of the two strange-charmed mesons $D_{s0}(2317)$ and
$D_{s1}(2460)$ with spin-parity $0^+$ and $1^+$ respectively has
triggered hot  debate on their nature and under-structures
 \cite{exp03}. There  have
been a lot of explanations for  their nature, for example,
conventional $c\bar{s}$ states \cite{2quark,Dai2SR}, two-meson
molecular states \cite{2meson}, $D-K$ mixing states \cite{mixing}
and four-quark states \cite{4quark,Nielsen4SR}, etc. The mass of the
$D_{s0}(2317)$
 is significantly lower than the values of the $0^+$ state mass  from the quark models   and
lattice simulations \cite{QuarkLattice}. Those two states
$D_{s0}(2317)$ and $D_{s1}(2460)$ lie just below the $D K$ and
$D^\ast K$ threshold respectively which are analogous to  the
situation that the scalar mesons  $a_0(980)$ and $f_0(980)$ lie just
below the $K\bar{K}$ threshold and  couple strongly to the nearby
channels. If we take the scalar mesons $a_0(980)$ and $f_0(980)$ as
four-quark states with the
 constituents  of scalar diquark-antidiquark  sub-structures, the
masses of the scalar nonet mesons below  $1GeV$ can be naturally
explained. The mechanism responsible for the low-mass charmed scalar
meson may be the same as the light scalar  nonet mesons, the
$f_{0}(600)$, $f_{0}(980)$, $a_{0}(980)$  and $K^{\ast}_{0}(800)$
\cite{ReviewScalar,WangScalar05}. The one-gluon exchange force and
the instanton induced force  lead to significant attractions between
the quarks in the $0^+$ diquark channels, we can take the scalar
diquark and antidiquark as the basic constituents  in constructing
the interpolating current \cite{GluonInstanton}, furthermore, in the
color superconductivity theory, the attractive interactions in this
channel lead to the formulation of nonzero condensates and the
breaking of both the color and flavor $SU(3)$ symmetries for the
light flavors \cite{ReviewColor}.

In this article, we take the point of view that the charmed scalar
meson $D_{s0}(2317)$ be a tetraquark state consist of scalar diquark
and antidiquark, and devote to calculate its mass in the heavy quark
limit with the QCD sum rules \cite{SVZ}. The masses of the ground
state and lowest excited state heavy mesons have been studied with
QCD sum rules in heavy quark effective theory via the $1/m_Q$
expansion \cite{HeavySR, HeavySR2}.

The article is arranged as follows:  we derive the QCD sum rules for
the bound energy $\bar{\Lambda}$ of the $D_{s0}(2317)$ in the heavy
quark limit in section II; in section III, numerical results and
discussions; section VI is reserved for conclusion.

\section{QCD sum rules for the $D_{s0}(2317)$ in the heavy quark limit}

In the following, we write down  the two-point correlation function
$\Pi $ in the framework of the QCD sum rules approach
\cite{Nielsen4SR},
\begin{eqnarray}
\Pi &=&i\int d^4x e^{ik \cdot x} \langle
0|T\{J(x)J^+(0)\}|0\rangle,  \\
J(x)&=&{\epsilon^{kij}\epsilon^{kmn}\over\sqrt{2}}\left\{u_i^T(x)C
\gamma_5c_j(x)\bar{u}_m(x)\gamma_5C\bar{s}_n^T(x)+d_i^T(x)C
\gamma_5c_j(x)\bar{d}_m(x)\gamma_5C\bar{s}_n^T(x)\right\} , \nonumber \\
\end{eqnarray}
here the $i,j,k,m,n $ are color indices and the $C$ is the charge
conjugation matrix. In the heavy quark limit, the $c$ quark field
can be approximated by the static heavy quark field $h_v(x)$ with
the propagator,
\begin{eqnarray}
\langle0| T[h_v(x)\bar{h}_v(0)]|0 \rangle=\frac{1+\!\not\!
v}{2}\int_0^{\infty}\delta(x-vt)dt,
\end{eqnarray}
here the $v_\mu$ is a four-vector with $v^2=1$.  The calculation of
the operator product expansion can be performed in the coordinate
space, and does not need the mixed picture both in coordinate and
momentum spaces \cite{Nielsen4SR}, in the heavy quark limit, the
calculation can be greatly facilitated.

According to the basic assumption of current-hadron duality in the
QCD sum rules approach \cite{SVZ}, we insert  a complete series of
intermediate states satisfying the unitarity principle with the same
quantum numbers as the current operator $J(0)$
 into the correlation function in
Eq.(1)  to obtain the hadronic representation. After isolating the
pole term  of the lowest $D_{s0}(2317)$  state, we obtain the
following result in the heavy quark limit,
\begin{eqnarray}
\Pi &=&\frac{F^2}{\bar{\Lambda}- \omega}+\cdots , \nonumber\\
\langle0|J(0)|D_{s0}\rangle&=&\sqrt{2}F, \nonumber \\
\bar{\Lambda}&=&limit_{m_c\rightarrow\infty}m_{D_{s0}}-m_c,
\end{eqnarray}
here $\omega=v\cdot k$ and $\bar{\Lambda}$ is the bound  energy.

 We perform operator product expansion up to the vacuum condensates
of dimension-9 to obtain the correlation function $\Pi $ at the
level of quark-gluon degrees of freedom. Once  the analytical
results are obtained,
  then we can take the current-hadron duality  below the threshold
$\omega_c$ and perform the Borel transformation with respect to the
variable $\omega$, finally we obtain  the following sum rule,
\begin{eqnarray}
F^2 e^{-\frac{\bar{\Lambda}}{T}}&=&\int_{m_s}^{\omega_c}d\omega
e^{-\frac{\omega}{T}}\left\{
\frac{\omega^8}{3360\pi^{6}}-\frac{\langle\bar{q}q\rangle
\omega^5}{60\pi^4}
-\frac{m_s\left(2\langle\bar{q}q\rangle-\langle\bar{s}s\rangle\right)\omega^4
}{48\pi^4} \right.\nonumber\\
&&\left.+\frac{\omega^4}{384\pi^4}\langle\frac{\alpha_sGG}{\pi}\rangle
+\frac{\langle\bar{q}g_s\sigma Gq\rangle
\omega^3}{48\pi^4}+\frac{m_s\left(3\langle\bar{q}g_s\sigma
Gq\rangle-\langle\bar{s}g_s\sigma Gs\rangle\right)
\omega^2}{96\pi^4} \right.\nonumber\\
&&\left. +\frac{\langle\bar{q}q\rangle\langle\bar{s}s\rangle
\omega^2}{6\pi^2}+\frac{m_s\left(2\langle\bar{q}q\rangle^2-\langle\bar{q}q\rangle\langle\bar{s}s\rangle\right)\omega}{12\pi^2}
-\frac{\langle\bar{q}q\rangle^2\langle\bar{s}s\rangle}{18}\delta(\omega)\right\},
\end{eqnarray}
here the $T$ is the Borel parameter.
Differentiate the above sum
rule with respect to the variable $\frac{1}{T}$, then eliminate the
quantity $F$, we obtain
\begin{eqnarray}
\bar{\Lambda}&=&\int_{m_s}^{\omega_c}d\omega
e^{-\frac{\omega}{T}}\left\{
\frac{\omega^9}{3360\pi^{6}}-\frac{\langle\bar{q}q\rangle
\omega^6}{60\pi^4}
-\frac{m_s\left(2\langle\bar{q}q\rangle-\langle\bar{s}s\rangle\right)\omega^5
}{48\pi^4} \right.\nonumber\\
&&\left.+\frac{\omega^5}{384\pi^4}\langle\frac{\alpha_sGG}{\pi}\rangle
+\frac{\langle\bar{q}g_s\sigma Gq\rangle
\omega^4}{48\pi^4}+\frac{m_s\left(3\langle\bar{q}g_s\sigma
Gq\rangle-\langle\bar{s}g_s\sigma Gs\rangle\right)
\omega^3}{96\pi^4} \right.\nonumber\\
&&\left. +\frac{\langle\bar{q}q\rangle\langle\bar{s}s\rangle
\omega^3}{6\pi^2}+\frac{m_s\left(2\langle\bar{q}q\rangle^2-\langle\bar{q}q\rangle\langle\bar{s}s\rangle\right)\omega^2}{12\pi^2}
\right\}/ \int_{m_s}^{\omega_c}d\omega
e^{-\frac{\omega}{T}}\nonumber\\
&&\left\{ \frac{\omega^8}{3360\pi^{6}}-\frac{\langle\bar{q}q\rangle
\omega^5}{60\pi^4}
-\frac{m_s\left(2\langle\bar{q}q\rangle-\langle\bar{s}s\rangle\right)\omega^4
}{48\pi^4} +\frac{\omega^4}{384\pi^4}\langle\frac{\alpha_sGG}{\pi}\rangle\right.\nonumber\\
&&\left. +\frac{\langle\bar{q}g_s\sigma Gq\rangle
\omega^3}{48\pi^4}+\frac{m_s\left(3\langle\bar{q}g_s\sigma
Gq\rangle-\langle\bar{s}g_s\sigma Gs\rangle\right)
\omega^2}{96\pi^4} \right.\nonumber\\
&&\left. +\frac{\langle\bar{q}q\rangle\langle\bar{s}s\rangle
\omega^2}{6\pi^2}+\frac{m_s\left(2\langle\bar{q}q\rangle^2-\langle\bar{q}q\rangle\langle\bar{s}s\rangle\right)\omega}{12\pi^2}
-\frac{\langle\bar{q}q\rangle^2\langle\bar{s}s\rangle}{18}\delta(\omega)\right\}.
\end{eqnarray}
It is easy to integrate over  the variable  $\omega$, we prefer this
formulation for simplicity.
\section{Numerical results and discussions}
 The parameters for the condensates are chosen to
be the standard values at the energy scale $\mu=1GeV$, although
there are some suggestions for updating those values, for reviews,
one can consult Ref.\cite{Update}. $\langle \bar{s}s \rangle=0.8
\langle \bar{q}q \rangle$, $\langle \bar{q}q \rangle=\langle
\bar{u}u \rangle=\langle \bar{d}d \rangle=-(240MeV)^3$, $\langle
\bar{q}g_s\sigma G q \rangle=m_0^2\langle \bar{q}q \rangle$,
$\langle \bar{s}g_s\sigma G s \rangle=m_0^2\langle \bar{s}s
\rangle$, $m_0^2=0.8GeV^2$,
 $\langle \frac{\alpha_s GG}{\pi}
\rangle=(0.33GeV)^4$ and
 $m_u=m_d=0$. Small variations of those condensates will not
 lead to large  changes for   the numerical  values and impair the predictive ability,
  we can neglect  the uncertainties for the vacuum condensates for simplicity.
  The mass of the $s$ quark from the Particle Data Group  is about
 $m_s(\mu=2GeV)=(80-155)MeV$ \cite{PDG}, the values (listed in the first article of Ref.\cite{Update}) from
 the QCD sum rules, lattice QCD and $\tau$ decays vary in a large range,
 $m_s(\mu=1GeV)\approx(117-203)MeV$,
here we take the average values from  lattice QCD,
$m_s(\mu=1GeV)=(140\pm10)MeV$. The variations of the $m_s$ about
$20MeV$ can only lead to tiny  changes for the final result, the
uncertainties can be safely neglected, we take the value
$m_s(\mu=1GeV)=140MeV$ in numerical calculation for simplicity. The
values of the mass of the $c$ quark are $m_c=(1.0-1.4)GeV$ at the
energy scale   $\mu=2GeV$ from
 the Particle Data Group  \cite{PDG}. The average values (listed in the first article of Ref.\cite{Update})
 at the energy scale $\mu=m_c$ are
 $m_c(m_c)=(1.3\pm0.1)GeV$ from the QCD sum rules and lattice QCD.
 We take the values $m_c(m_c)=(1.3\pm0.1)GeV$ and evolve them  to
 the  energy scale $\mu=1GeV$ with the renormalization group equation,
 $m_c(\mu=1GeV)=(1.4\pm 0.1)GeV$.

In the following, we discuss the  criterion  for selecting the
threshold parameter $s_0$ (or $\omega_c$) and Borel parameter $M_B$
(or $T$) in the QCD sum rules dealing with  the multiquark states.
The QCD sum rules have been extensively applied to the hadronic
physics and given a lot of successful  descriptions
\cite{SVZ,Update}. For the conventional (two-quark) mesons and
(three-quark) baryons, the hadronic  spectral densities are
experimentally well known, the separations between the ground state
and excited states are large enough, the "single-pole $+$ continuum
states" model works well in representing the phenomenological
spectral densities. In the phenomenological analysis, the continuum
states can be approximated by the contributions from the  asymptotic
quarks and gluons, and the single-pole dominance condition can be
well satisfied,
\begin{eqnarray}
\int_{s_0}^{\infty}\rho_{pert}e^{-\frac{s}{M_B^2}}ds <
\int^{s_0}_{0}(\rho_{pert}+\rho_{cond})e^{-\frac{s}{M_B^2}}ds
\end{eqnarray}
for the conventional QCD sum rules, and
\begin{eqnarray}
\int_{\omega_c}^{\infty}\rho_{pert}e^{-\frac{\omega}{T}}d\omega <
\int^{\omega_c}_{0}(\rho_{pert}+\rho_{cond})e^{-\frac{\omega}{T}}d\omega
\end{eqnarray}
for the QCD sum rules in the heavy quark limit,
 here the $\rho_{pert}$ and
$\rho_{cond}$ stand for the contributions from the perturbative and
non-perturbative part of the spectral density respectively. From the
 conditions in Eqs.(7-8), we can obtain the maximal value for the Borel parameter
$M_B^{max}$ (or $T_{max}$), exceed this value, the single-pole
dominance will be spoiled. On the other hand, the Borel parameter
must be chosen large enough to warrant the convergence of the
operator product expansion and contributions from the high dimension
vacuum condensates which are poorly known are of minor importance,
the minimal value for the Borel parameter $M_B^{min}$ (or $T_{min}$)
can be determined.

For the conventional (two-quark) mesons and (three-quark) baryons,
the Borel window $M_B^{max}-M_B^{min}$ (or $T_{max}-T_{min}$) is
rather large and the reliable QCD sum rules can be obtained.
However, for the multiquark states i.e. tetraquark states,
pentaquark states, etc, the spectral densities $\rho\sim s^m$ with
$m$ is larger than the corresponding ones for the conventional
hadrons,   the integral $\int_0^{\infty} s^m
\exp\left(-\frac{s}{M_B^2}\right) ds$ (or $\int_0^{\infty} \omega^m
\exp\left(-\frac{\omega}{T}\right) d\omega$) converges more slowly
\cite{Narison04}. If one do not want to release the conditions  in
Eqs.(7-8), we have to either postpone the threshold parameter $s_0$
(or $\omega_c$) to very large values or choose very small values for
the Borel Parameter $M_B^{max}$ (or $T_{max}$). With large values
for the threshold parameter $s_0$ (or $\omega_c$), for example, $s_0
\gg M_{gr}^2$, here the $gr$ stands for the ground state, the
contributions from the excited states are already included in if
there are really some ones, the single-pole approximation for the
spectral densities is spoiled; on the other hand, with very small
values for the Borel parameter $M_B^{max}$ (or $T_{max}$), the
operator product expansion is broken down, and the Borel window
shrinks to zero and beyond. This may lead to the pessimistic opinion
that  the QCD sum rules can not be successfully applied to the
multiquark states, the sum rules concerning the tetraquark states
and pentaquark states should be rejected, however, we are
optimistical  participators for the QCD sum rules and take the point
of view that the QCD sum rules can be successfully  applied to the
multiquark states, and one should  resort to  the "multi-pole $+$
continuum states" to approximate the phenomenological spectral
densities. The onset of the continuum states  is not abrupt,   the
ground state, the first excited state, the second excited state,
etc, the continuum states appear sequentially; the excited states
may be loose  bound states and have large widths. The threshold
parameter $s_0$ (or $\omega_c$) is postponed to large value, at that
energy scale, the spectral densities can be well approximated by the
contributions from the asymptotic quarks and gluons, and of minor
importance for the sum rules.

The present experimental knowledge about the phenomenological
hadronic spectral densities for the tetraquark states is  rather
vague, even the existence of the tetraquark states is not confirmed
with confidence, and no knowledge about either there are high
excited states or not. In this article,  the following criteria are
taken. We choose the suitable values for the Borel parameter $M_B$
(or $T$), on the one hand the minimal values $M_B^{min}$ (or
$T_{min}$) are large enough to warrant the convergence of the
operator product expansion, on the other hand the maximal values
$M_B^{max}$ ($T_{max}$) are small enough to suppress the
contributions from the high excited states and continuum states i.e.
 we choose the naive analysis $M_B^{max}< M_{gr}$ (or
$T_{max}<\bar{\Lambda}$), furthermore, there exist a Borel platform
which is   insensitive to the variations of the Borel parameter. For
the hadronic spectral density, the more phenomenological analysis is
preferred, we approximate the spectral densities with the
contribution from the single-pole term, the threshold parameter
$s_0$ (or $\omega_c$) is taken slightly above the ground state mass,
$s_0>M_{gr}^2$ (or $\omega_c>\bar{\Lambda}$), to subtract the
contributions from the excited states and continuum states. One may
reject taking the values from the more phenomenological analysis as
quantitatively reliable, the results are qualitative at least.

 In the heavy quark limit $\sqrt{s_0}\sim m_c+\omega_c  $,
 the threshold parameter $\omega_c$ is chosen to  vary between $(1.3-1.5)
 GeV$.  The values are
reasonable  for the scalar meson  $D_{s0}(2317)$ with narrow width,
  $[ m_c+(1.3-1.5)GeV] \geq(2.6-2.8)GeV > M_{D_{s0}}+\Gamma_{D_{s0}}$,
  the contributions from the $D_{s0}(2317)$ can be
  correctly taken into account. For the values $\omega_c\geq 1.3GeV$,
  the contributions from the perturbative term and
the linear quark condensate terms proportional to the
$\langle\bar{q}q\rangle$, $\langle\bar{s}s\rangle$ are dominating
i.e. $ > 60\%$ with the minimal Borel parameter $T_{min}\geq0.5GeV$,
in this region, we can warrant the convergence of the operator
product expansion.
  For the intervals, $T=(0.5-1.1)GeV$ and $\omega_c=(1.3-1.5)GeV$, the main
contributions to the bound energy $\bar{\Lambda}$ come from the
linear  quark condensates terms proportional to the $\langle
\bar{q}q\rangle $ and $\langle \bar{s}s\rangle $, about $(60-70)\%$,
the perturbative contributions are suppressed by the large numerical
denominator and of minor importance, about $(10-15)\%$, which is
significantly in contrary to the ordinary QCD sum rules for the
conventional  mesons and baryons where the main contributions come
from the perturbative terms. Furthermore, the contributions from the
gluon condensates are of minor importance due to
 the large numerical denominator \cite{6quark}.  From the Fig.1, we can see that  the
predicted bound energy $\bar{\Lambda}$ is almost independent  on the
Borel parameter $T$ in the region $ 0.5GeV \leq T \leq 1.1GeV^2$. If
we restrict the values of the Borel parameter $T$ to the naive
analysis $T_{max}<{\bar{\Lambda}}$, the predicted values are
$\bar{\Lambda}=(0.8-1.0)GeV$ with $\omega_c=(1.3-1.5)GeV$. For
$T_{max}=0.8$, the contributions from the perturbative term and the
linear quark condensate terms proportional to the
$\langle\bar{q}q\rangle$, $\langle\bar{s}s\rangle$ are larger than
$85\%$ with the threshold parameter $\omega_c>1.5GeV$, the high
dimensional condensates (non-perturbative terms) are greatly
suppressed, this may be an  indication the onset of the  continuum
states.
 For $\omega_c=(1.3-1.5)GeV$ and $T=(0.5-0.8)GeV$,
\begin{eqnarray}
\bar{\Lambda}&=&(0.8-1.0)GeV ,  \\
m_{D_{s0}}&=&[m_c+(0.8-1.0)]GeV ,\nonumber \\
&=&[(1.3-1.5)+(0.8-1.0)]GeV ,\nonumber \\
&=& (2.1-2.5)GeV , \nonumber \\
&=&(2.3\pm0.2)GeV .
\end{eqnarray}
Comparing with the experimental data, the tetraquark configuration
gives reasonable values for the mass of the scalar meson
$D_{s0}(2317)$; there must be some tetraquark component in the
charmed scalar meson $D_{s0}(2317)$.
\begin{figure}
 \centering
 \includegraphics[totalheight=7cm]{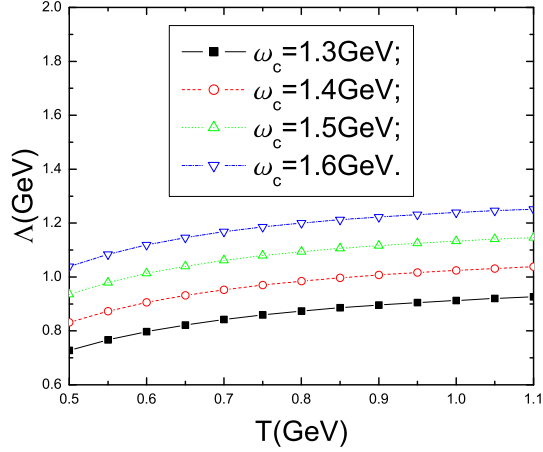}
 \caption{The   $\bar{\Lambda}$ with the Borel parameter $T$ . }
\end{figure}

\section{Conclusion}
In this article, we take the point of view that the charmed scalar
meson $D_{s0}(2317)$ be a tetraquark state and devote to calculate
its mass within the framework of the QCD sum rule approach in the
heavy quark limit. The numerical values for the mass of the
$D_{s0}(2317)$ are consistent with the experimental data. There must
be some tetraquark component in the scalar meson $D_{s0}(2317)$.

\section*{Acknowledgments}
This  work is supported by National Natural Science Foundation,
Grant Number 10405009,  and Key Program Foundation of NCEPU. The
authors are indebted to Dr. J.He (IHEP), Dr. X.B.Huang (PKU) and Dr.
L.Li (GSCAS) for numerous help, without them, the work would not be
finished.

\end{document}